\documentclass[conference]{IEEEtran}
\IEEEoverridecommandlockouts
\usepackage{cite}
\usepackage{amsmath,amssymb,amsfonts}
\usepackage{algorithmic}
\usepackage{algorithm}
\usepackage{graphicx}
\usepackage{textcomp}
\usepackage{xcolor}
\newcommand{\rev}[1]{\textcolor{black}{#1}}
\usepackage{etoolbox}
\usepackage[hidelinks]{hyperref}
\def\BibTeX{{\rm B\kern-.05em{\sc i\kern-.025em b}\kern-.08em
    T\kern-.1667em\lower.7ex\hbox{E}\kern-.125emX}}

\allowdisplaybreaks

\usepackage[justification=justified,font=footnotesize,skip=0pt]{caption}

\begin{document}

\title{Risk Assessment of Fair Multi-Period Dynamic Operating Envelopes under Uncertainty}

\author{Pedro Salom{\~a}o Quessongo$^\dagger$, Daniel Gebbran$^\dagger$, \textit{Member, IEEE}, Wilhiam de Carvalho$^\ddagger$, \\Clodomiro Unsihuay-Vila$^\dagger$, \textit{Member, IEEE}%
\vspace{-0.5cm}
\thanks{This work was financially supported by the Brazilian National Council for Scientific and Technological Development --- CNPq, under project grant 447271/2024-5.
$^\dagger$Department of Electrical Engineering, Federal University of Paran{\'a} (UFPR), Curitiba, Brazil.
$^\ddagger$Luxembourg Institute of Science and Technology (LIST), Esch-sur-Alzette, Luxembourg.
Email: \{\texttt{pedro.quessongo}, \texttt{daniel.gebbran}, \texttt{clodomiro.vila}\}@ufpr.br, \texttt{\{wilhiam.carvalho\}}@list.lu.}
}

\maketitle

\begin{abstract}

Dynamic operating envelopes (DOEs) can allocate time-varying export capacity to distributed energy resources (DERs), and while they are not intended to make operations immune to voltage violations, they should guide operations to mostly remain inside expected values under changing regimes and residual forecast errors. However, although recent literature proposed many different techniques for implementing DOEs, there is sparse discussion of analyzing and quantifying the risk of using more or less conservative DOEs under uncertain operating regimes. Having this gap in mind, this paper develops a risk-oriented extension of a previously established fair distributed multi-period optimal power flow (OPF) framework, where fair or technical DOEs are fixed as first-stage decisions and the OPF leverages battery energy storage and renewable curtailment as recourse under regime uncertainties. An independent nonlinear AC power flow then computes a resulting ensemble, which is summarized by (i) voltage-violation probability, severity, and a persistence metric for consecutive out-of-band hours, (ii) curtailment and Jain and Gini indices, and (iii) model deviations. Different probability situations are assessed, ranging from uniform, sunny-heavy, night-peak-heavy, and adversarial, which separate the envelope design from the assumed operating distribution. An optional outer loop allows for fine-tuning a fairness–efficiency parameter under explicit violation and persistence budgets. The combination of the outer loop and the overall framework, which quantifies how often and how severely operating limits are exceeded, creates an auditable methodology mixing equitable export allocation, distributed recourse, and AC risk assessment for active distribution network operators, \rev{and this framework is assessed using up to 1000 Monte Carlo scenarios per mixture and DOE design}.

\end{abstract}


\section{Introduction}

High penetrations of rooftop photovoltaic (PV), wind generation, and battery energy storage systems (BESS) expose distribution feeders to reverse power flow, voltage rise, congestion, and time-dependent scarcity of export capacity~\cite{REN21_2025}. Dynamic operating envelopes (DOEs) replace uniform static limits with network-aware, time-varying permissions~\cite{lankeshwara2025}, which often use optimal power flow (OPF) methods as an underlying tool for computing such operating envelopes in a dynamic, network-aware fashion~\cite{wickramasinghe2025}.

In particular, branch-flow OPF models provide an efficient basis for radial distribution operation~\cite{baran1989,low2014}, while multi-period distributed OPF extends this basis to storage and temporal coupling~\cite{pinto2020}, and the use of the alternating direction method of multipliers (ADMM) is often preferred for enabling such regional coordination, relying only on limited information exchange on the shared interfaces.
Furthermore, the allocation of fair DOEs has been discussed as a way to promote more equitable access, which prevents favoring prosumers at electrically strong buses. Authors have proposed the use of proportional rules~\cite{petrou2020} and uncertainty-aware formulations~\cite{yi2022}.

No recent work, however, addressed these different multi-period and fairness aspects in a single formulation. As such, the authors' preceding work introduced a framework to compute DOEs with fairness across a multi-period problem, applying a cumulative proportional-fair allocation on top of technical DOEs, solving six representative operating regimes with BESS recourse in a distributed OPF solved using regional ADMM, and validating each final dispatch with AC power flow~\cite{quessongo2026}. The scenario-based results establish feasibility at the selected operating points, but they do not quantify how frequently limits are exceeded when the regimes occur with different probabilities or when forecasts contain residual errors.This is particularly important having in mind that DOEs do not intend to make operations immune to large violations (else, they would be far too conservative); still, operators need to know how often they would incur deviations given the selected parameters when deriving DOEs. This gap motivates the present work, and the main contributions can be summarized as:
\begin{itemize}
    \setlength{\itemsep}{2pt}
    \renewcommand{\labelitemi}{$\bullet$}
    \item The use of a regime-mixture uncertainty model built on six interpretable stress regimes, enforcing proportional equity, which are then solved and validated as per~\cite{quessongo2026};
    \item The establishment of an AC ensemble protocol to report violation (i)~probability, severity, and persistence, (ii)~curtailment and Jain and Gini indices, allowing for a same-distribution comparison of fair and technical DOEs, and (iii)~model deviations;
    \item The introduction of a computationally efficient lightweight outer loop that tunes fairness--efficiency parameters, having in mind the explicitly derived voltage-risk budgets.
\end{itemize}

\subsection{Related Work}
Research on DOEs spans technical envelope computation, equitable allocation, uncertainty-aware design, and probabilistic validation, as reviewed comprehensively in~\cite{lankeshwara2025,wickramasinghe2025}. Existing studies have developed Monte Carlo and nonlinear power-flow validation~\cite{ge2026}, weighted proportional fairness~\cite{petrou2020}, chance-constrained fair DOEs~\cite{yi2022}, separate technical and equitable allocation stages~\cite{alam2024}, bargaining-based allocation~\cite{jiang2025}, and stochastic or robust envelope construction~\cite{liu2024,koirala2024,decarvalho2026sest}. CVaR has also been applied to risk-aware DSO--microgrid coordination~\cite{ge2026}. Unlike these approaches, the present work fixes previously computed fair or technical DOEs before sampling and evaluates them using paired multiperiod AC ensembles with scenario-dependent BESS, curtailment, and emergency recourse. Its contribution is the combined assessment of cumulative fairness, CVaR, nonlinear AC feasibility, and temporal persistence.

\section{Deterministic DOE Model}
\label{sec:deterministic}

This section restates the inherited multiperiod DOE/OPF model in self-contained form, so that the risk layer that follows can be read without consulting~\cite{quessongo2026} for every symbol. Operationally, the workflow is staged: technical envelopes are computed first; fair envelopes optionally redistribute that capacity under a budgeted equity criterion; for each uncertain realization the chosen envelopes are held fixed while BESS, curtailment, and (if needed) emergency demand response act as recourse; finally, the consolidated injections are validated by nonlinear AC power flow.

\subsection{Network and Recourse Dispatch}
Let $\mathcal{N}$, $\mathcal{E}$, $\mathcal{T}$, $\mathcal{P}$, and $\mathcal{B}$ denote the sets of buses, radial branches, time periods, renewable prosumers, and BESS units, respectively. \rev{To make the inherited aggregation explicit, $M_{in}=1$ if prosumer $i$ is connected to bus $n$ and zero otherwise, while $J_{ig}=1$ if DER $g$ belongs to prosumer $i$ and zero otherwise. Hence bus-$n$ DER injection aggregates $\sum_i M_{in}\sum_g J_{ig}p_{g,t}$, and prosumer-level energy quantities aggregate the same devices over $\mathcal{T}$.} For each branch $(i,j)\in\mathcal{E}$, the active and reactive flows are $P_{ij,t}$ and $Q_{ij,t}$, and $v_{i,t}$ is the squared voltage magnitude at bus~$i$. We adopt the lossless LinDistFlow equations
\begin{subequations}\label{eq:lindist}
\begin{align}
P_{ij,t} &= \sum_{k\in\mathrm{child}(j)} P_{jk,t}
+ p^{L}_{j,t} - p^{R}_{j,t} + p^{ch}_{j,t} - p^{dis}_{j,t}, \label{eq:Pflow}\\
Q_{ij,t} &= \sum_{k\in\mathrm{child}(j)} Q_{jk,t}
+ q^{L}_{j,t} - q^{DER}_{j,t}, \label{eq:Qflow}\\
v_{j,t} &= v_{i,t} - 2\big(r_{ij} P_{ij,t} + x_{ij} Q_{ij,t}\big). \label{eq:vdrop}
\end{align}
\end{subequations}
Voltage and thermal limits, renewable availability/curtailment, and the fixed-envelope export cap associated with~\eqref{eq:lindist} are
\begin{subequations}\label{eq:limits}
\begin{align}
\underline{v} &\leq v_{i,t} \leq \overline{v},
\qquad P_{ij,t}^{2}+Q_{ij,t}^{2}\leq S_{ij,\max}^{2}, \label{eq:vtherm}\\
0 &\leq p^{R}_{i,t} \leq p^{R,\mathrm{avail}}_{i,t},
\qquad p^{\mathrm{curt}}_{i,t}=p^{R,\mathrm{avail}}_{i,t}-p^{R}_{i,t}, \label{eq:ren}\\
p^{R}_{i,t} &\leq d_{i,t}, \label{eq:doe_cap}
\end{align}
\end{subequations}
where $d_{i,t}$ is the published (technical or fair) DOE. The intertemporal BESS balance, together with its power/energy limits and terminal-state condition, follows the multiperiod storage treatment in~\cite{lankeshwara2025} and the implementation in~\cite{quessongo2026}:
\begin{equation}\label{eq:bess}
e_{b,t+1}
=
e_{b,t}+\eta^{ch}p^{ch}_{b,t}\Delta t
-\frac{p^{dis}_{b,t}\Delta t}{\eta^{dis}}.
\end{equation}
The inherited recourse cost in~\cite{quessongo2026} includes grid exchange, curtailment, BESS degradation, and a prohibitive emergency-flexibility penalty; positive emergency demand response therefore denotes restored, not strict, feasibility.

\subsection{Technical and Fair DOE Families}
The technical envelope $d^{\mathrm{tech}}$ is obtained from a network-constrained export-capacity problem. The fair envelope $d^{\mathrm{fair}}$ redistributes a constrained aggregate budget while respecting
\begin{equation}\label{eq:fair_box}
0 \leq d^{\mathrm{fair}}_{i,t} \leq d^{\mathrm{tech}}_{i,t}.
\end{equation}
For each prosumer $i$, \rev{define $E_i^{\mathrm{avail}}=\sum_gJ_{ig}\sum_t p_{g,t}^{R,\mathrm{avail}}\Delta t$, $E_i^{\mathrm{curt}}=\sum_gJ_{ig}\sum_t p_{g,t}^{\mathrm{curt}}\Delta t$, and $E_i^{\mathrm{acc}}=E_i^{\mathrm{avail}}-E_i^{\mathrm{curt}}$.} The acceptance ratio is
\begin{equation}\label{eq:eta}
\eta_i = \frac{E^{\mathrm{acc}}_i}{E^{\mathrm{avail}}_i + \kappa},
\qquad \kappa=10^{-6}\max_{j\in\mathcal{P}}E_j^{\mathrm{avail}},
\end{equation}
\rev{where $\kappa>0$ is a numerical regularizer that prevents division by zero without materially changing ratios for nonzero availability. For the curtailment-based Gini index, $\chi_i=E_i^{\mathrm{curt}}/(E_i^{\mathrm{avail}}+\kappa)$.}
Ex-post acceptance equity is then measured by Jain's fairness index (JFI),
\begin{equation}\label{eq:jfi}
\mathrm{JFI}
=
\frac{\big(\sum_{i\in\mathcal{P}}\eta_i\big)^{2}}
{|\mathcal{P}|\sum_{i\in\mathcal{P}}\eta_i^{2}},
\end{equation}
and the Gini coefficient of $\{\chi_i\}$ detects concentration that may be hidden by a high acceptance-based Jain index:
\begin{equation}\label{eq:gini}
G
=
\frac{\sum_{i=1}^{|\mathcal{P}|}\sum_{j=1}^{|\mathcal{P}|}|\chi_i-\chi_j|}
{2|\mathcal{P}|\sum_{i=1}^{|\mathcal{P}|}\chi_i}.
\end{equation}
A one-parameter family of fair envelopes $d^{\mathrm{fair}}(\theta)$ can be defined, where $\theta$ represents the aggregate-budget scaling factor $\beta$, the allowed fairness--efficiency gap $\delta$, or a uniform/network-sensitive derating factor. The DOE is fixed before uncertainty is sampled; only BESS, curtailment, and emergency flexibility are treated as recourse decisions.

\subsection{Stagewise Objective Functions}
The framework contains three optimization stages and their objectives must not be conflated. Let $C^{\mathrm{op}}(x,\omega)$ denote the grid-energy cost minus export revenue plus renewable-curtailment, BESS-degradation, and emergency-flexibility penalties.

The technical stage maximizes feasible envelope capacity through the equivalent minimization
\begin{equation}\label{eq:tech_obj}
\min_{d^{\mathrm{tech}}} \; -\sum_{i\in\mathcal{P}}\sum_{t\in\mathcal{T}} d^{\mathrm{tech}}_{i,t}
\end{equation}
subject to power-flow, DER, BESS, voltage, thermal, and substation constraints. After applying
\begin{equation}\label{eq:Dcap}
D^{\mathrm{cap}}_{t}=\beta_{t}\sum_{i\in\mathcal{P}} d^{\mathrm{tech}}_{i,t},
\end{equation}
the fairness stage determines $d^{\mathrm{fair}}$ and the associated nominal dispatch by
\begin{subequations}\label{eq:fair}
\begin{align}
\min_{x,\,d^{\mathrm{fair}},\,\Gamma}\;
&\Gamma+\lambda_{c}E^{\mathrm{curt}}
-\lambda_{u}\sum_{i,t}d^{\mathrm{fair}}_{i,t}
+\lambda_{0}\big\|d^{\mathrm{fair}}-d^{0}\big\|_{2}^{2}
\label{eq:fair_obj}\\
\text{s.t.}\quad
&0\leq d^{\mathrm{fair}}_{i,t}\leq d^{\mathrm{tech}}_{i,t},
\quad \sum_{i}d^{\mathrm{fair}}_{i,t}\leq D^{\mathrm{cap}}_{t},
\label{eq:fair_budget}\\
&E^{\mathrm{curt}}_{i}\leq\Gamma\big(E^{\mathrm{avail}}_{i}+\kappa\big),\nonumber\\[-1pt]
&E^{\mathrm{curt}}\leq E^{\mathrm{curt,tech}}+\delta E^{\mathrm{avail}}.
\label{eq:fair_gamma}
\end{align}
\end{subequations}
The small terms $\lambda_{c}$, $\lambda_{u}$, and $\lambda_{0}$ break ties without replacing the min--max fairness priority.

For each sampled realization, the DOE is fixed and the recourse stage solves
\begin{align}\label{eq:recourse}
\min_{x^{\omega}}\quad &C^{\mathrm{op}}(x^{\omega},\omega)\nonumber\\[-1pt]
&=
\sum_{t\in\mathcal{T}}\Big(
c^{\mathrm{imp}}_{t}p^{\mathrm{imp}}_{t}
-c^{\mathrm{exp}}p^{\mathrm{exp}}_{t}
+c^{\mathrm{curt}}\sum_{i}p^{\mathrm{curt}}_{i,t} \nonumber\\
&\quad+c^{\mathrm{deg}}\sum_{b}(p^{ch}_{b,t}+p^{dis}_{b,t})\nonumber\\[-1pt]
&\quad+c^{\mathrm{DR}}\sum_{i}p^{\mathrm{DR}}_{i,t}
\Big)\Delta t.
\end{align}
The emergency coefficient $c^{\mathrm{DR}}$ is prohibitive. Therefore, $p^{\mathrm{DR}}_{i,t}>0$ identifies security restoration and is reported as strict infeasibility, not as normal recourse.

In the distributed implementation, each regional objective contains its local share of~\eqref{eq:recourse} plus the scaled ADMM consensus penalty. Formally, for region $k$,
\begin{equation}\label{eq:admm}
\min_{x^{\omega}_{k}}\;
C^{\mathrm{op}}_{k}(x^{\omega}_{k},\omega)
+\frac{\rho}{2}\big\|x^{\mathrm{boundary}}_{k}-z^{\omega}+u^{\omega}_{k}\big\|_{2}^{2},
\end{equation}
where $z^{\omega}$ is the global consensus variable and $u^{\omega}_{k}$ is the dual multiplier. ADMM changes the solution mechanism, not the centralized economic objective:
\begin{equation}\label{eq:op_sum}
C^{\mathrm{op}}(x^{\omega},\omega)=\sum_{k}C^{\mathrm{op}}_{k}(x^{\omega}_{k},\omega).
\end{equation}

\section{Regime-Mixture Uncertainty}
\label{sec:regimes}

\subsection{Discrete Regimes and Residual Error}
Let $\Omega=\{\omega_1,\ldots,\omega_6\}$ denote nominal, high-PV/low-load, high-wind, peak-load/low-renewable, night-peak, and mixed-congestion regimes. A categorical law $\pi$ assigns occurrence probabilities.

Conditional residual error $\xi|\omega$ perturbs load, PV, and wind trajectories while preserving the physical character of the parent regime. \rev{The implemented nominal model uses truncated, temporally correlated errors with $\sigma_L=3\%$, $\sigma_{PV}=8\%$, $\sigma_W=10\%$, temporal correlation $\rho=0.75$, and truncation at three standard deviations. Common random numbers pair fair and technical DOE evaluations. Tuning and test seeds are separated.} %($2026092801$ and $2026092802$, respectively). 
\rev{Sample sizes used later are $N{=}600$ for the developmental Uniform noise ablation (Table~\ref{tab:noise-ablation-results}) and $N{=}1000$ for the held-out Uniform/Adversarial tests at $\beta^\star=0.80$ (Table~\ref{tab:ensemble-results}); the adversarial fair/nominal ablation cell remains a preliminary $N{=}34$ entry.}
% \rev{The developmental Uniform ablation contains 600 successful samples per reported DOE/noise setting. The final held-out evaluation contains 1000 successful nominal-noise samples for each of the Uniform and Adversarial laws at $\beta^\star=0.80$. The earlier 34-sample adversarial fair/nominal entry remains preliminary and is not used for final robust-DOE claims.}

\begin{table}[t]
\caption{Regime-mixture laws used in the minimal experiment}
\label{tab:probability}
\centering
\begin{tabular}{lcccccc}
\hline
Law & $\omega_1$ & $\omega_2$ & $\omega_3$ & $\omega_4$ & $\omega_5$ & $\omega_6$ \\
\hline
Uniform      & .167 & .167 & .167 & .167 & .167 & .167 \\
Sunny-heavy  & .15  & .40  & .15  & .10  & .05  & .15  \\
Night-peak   & .15  & .08  & .07  & .15  & .40  & .15  \\
Adversarial  & .05  & .05  & .05  & .40  & .40  & .05  \\
\hline
\end{tabular}
\end{table}

Table~\ref{tab:probability} presents the probability assigned to each operating regime. The uniform law assigns equal probability to all regimes, whereas the sunny-heavy and night-peak laws emphasize $\omega_2$ and $\omega_5$, respectively. The adversarial law assigns $80\%$ of the total probability to the critical regimes $\omega_4$ and $\omega_5$, thereby assessing DOE performance under more stressful operating conditions. %Fig.~\ref{fig:regimes} visualizes the same mixture laws.

% \begin{figure}[t]
%     \centering
%     \includegraphics[width=0.90\linewidth]{fig.2.png}
%     \caption{Probability mass assigned to the six operating regimes by each experimental mixture law.}
%     \label{fig:regimes}
% \end{figure}

\subsection{Samplewise Evaluation Pipeline}
% For each sampled $(\omega,\xi)$, the selected DOE is fixed. The recourse DOPF is then solved by the existing regional ADMM implementation. For large ensembles, a local feasible dispatch heuristic may be substituted, but its optimality gap and failure rate must be disclosed.

For each sampled $(\omega,\xi)$, the selected DOE is fixed. The recourse DOPF is then solved by the existing regional ADMM implementation. If ADMM does not converge, the disclosed centralized recourse fallback is used and its share is reported.

The consolidated DER/BESS set-points are passed to an independent nonlinear AC power flow. Failed optimizations and failed AC power flows are counted and reported; silently discarding them would bias the risk estimate downward. The empirical bus--hour violation rate is
\begin{equation}\label{eq:pviol}
\hat{p}_{\mathrm{viol}}
=
\frac{1}{|\mathcal{M}|\,|\mathcal{N}|\,|\mathcal{T}|}
\sum_{m\in\mathcal{M}}\sum_{i\in\mathcal{N}}\sum_{t\in\mathcal{T}}
\mathbf{1}\big\{V^{\mathrm{AC},m}_{i,t}\notin[\underline{V},\overline{V}]\big\}.
\end{equation}
It complements, rather than replaces, scenario-level feasibility. This study reports failed-solve, emergency-recourse, and fallback shares; day-level bootstrap intervals are not tabulated.
% It complements, rather than replaces, scenario-level feasibility. A binomial confidence interval is reported by treating sampled day-level violation counts as clustered observations, or by nonparametric bootstrap at the day level.

% \begin{algorithm}[t]
%   \caption{Stochastic CVaR--Recourse AC Evaluation}
%   \label{algorithm:cvar-recourse}
%   \footnotesize
%   \begin{algorithmic}[1]
%     \STATE \textbf{Input:} fixed DOE $d$, $\pi$, $p(\xi|\omega)$, $M$, $\alpha$, and $H$.
%     \STATE Draw common-random-number samples $(\omega_m,\xi_m)$ and construct realized trajectories.
%     \STATE For each $m$, solve~\eqref{eq:recourse}; record recourse, fallback, convergence, and runtime.
%     \STATE Run independent AC power flow at the consolidated set-points and retain failed solves.
%     \STATE Compute exceedance, severity, persistence, curtailment, $\Delta V^{\rm AC}$, Jain, and Gini.
%     \STATE Aggregate means, maxima, CVaR, confidence bounds, and failure rates.
%   \end{algorithmic}
% \end{algorithm}

\begin{algorithm}[t]
  \caption{Stochastic CVaR--Recourse AC Evaluation}
  \label{algorithm:cvar-recourse}
  \footnotesize
  \begin{algorithmic}[1]
    \STATE \textbf{Input:} network and profiles; fixed DOE $d$; mixture $\pi$; conditional residual model $\xi|\omega$; sample size $M$; CVaR level $\alpha$; persistence horizon $H$.
    \STATE \textbf{Output:} samplewise AC states and recourse; $\hat{p}_{\mathrm{viol}}$, severity, $\hat{p}_{\mathrm{pers}}$, expected curtailment, $\mathrm{CVaR}_{\alpha}$, Jain/Gini, failure counts.
    \STATE Draw common-random-number samples $\{(\omega_m,\xi_m)\}$ from $\pi(\omega)p(\xi|\omega)$
    \STATE For $m=1,\ldots,M$, fix $d$ and construct the realized load, PV, and wind trajectories.
    % \STATE Solve recourse~\eqref{eq:recourse} by regional ADMM; if the disclosed fallback is triggered, apply the calibrated local feasible heuristic.
    \STATE Solve recourse~\eqref{eq:recourse} by regional ADMM; on nonconvergence, apply the disclosed centralized recourse fallback and record its share.
    \STATE Record BESS dispatch, curtailment, emergency DR, objective value, convergence status, and runtime.
    \STATE Fix the consolidated injections and run an independent nonlinear AC power flow; retain failed solves.
    \STATE Compute bus-hour exceedance, severity, $\Delta V^{\mathrm{AC}}$, run-length persistence, daily curtailment, and samplewise Jain/Gini.
    % \STATE Aggregate means, maxima, $\mathrm{CVaR}_{\alpha}$ via~\eqref{eq:cvar}, bootstrap confidence intervals, and failed-solve rates.
    \STATE Aggregate means, maxima, $\mathrm{CVaR}_{\alpha}$ via~\eqref{eq:cvar}, emergency recourse, fallback share, and failed-solve rates.
  \end{algorithmic}
\end{algorithm}

Algorithm~\ref{algorithm:cvar-recourse} evaluates a fixed DOE under uncertainty. For each sample, it solves the recourse dispatch using ADMM, validates the solution through nonlinear AC power flow, and calculates the resulting risk, curtailment, and fairness metrics.

\section{Risk, Equity, and Persistence Metrics}
\label{sec:metrics}

Voltage severity for sample $m$ is defined as
\begin{equation}\label{eq:severity}
e^{m}_{i,t}
=
\max\big\{\underline{V}-V^{\mathrm{AC},m}_{i,t},\;0,\;V^{\mathrm{AC},m}_{i,t}-\overline{V}\big\}.
\end{equation}
We report its sample mean, conditional mean given $e^{m}_{i,t}>0$, 95th percentile, and maximum. Model error is
\begin{equation}\label{eq:dVac}
\Delta V^{\mathrm{AC},m}_{i,t}
=
\big|V^{\mathrm{AC},m}_{i,t}-V^{\mathrm{opt},m}_{i,t}\big|,
\end{equation}
with both ensemble mean and maximum retained. Expected curtailment is the sample mean of $\sum_{i,t}p^{\mathrm{curt},m}_{i,t}\Delta t$. Expected Jain and Gini metrics are averages of samplewise indices, not indices computed after pooling all samples.

For a loss variable $L$, upper-tail risk is measured using~\cite{rockafellar2000}:
\begin{equation}\label{eq:cvar}
\mathrm{CVaR}_{\alpha}(L)
=
\min_{\eta\in\mathbb{R}}
\Big\{\eta+\tfrac{1}{1-\alpha}\,\mathbb{E}\big[(L-\eta)^{+}\big]\Big\}.
\end{equation}
Here $L$ may denote daily curtailment, operating cost, maximum voltage severity, or emergency recourse. CVaR complements the mean by emphasizing the worst $1-\alpha$ fraction of the regime mixture or Monte Carlo ensemble.

The run-length persistence metric distinguishes isolated spikes from sustained stress ($H=3\,\mathrm{h}$ primary; sensitivity $H\in\{2,4,6\}$):
\begin{subequations}\label{eq:pers}
\begin{align}
I_{\mathrm{pers}}(m,i)
&=
\mathbf{1}\Big\{\max_{t}\mathrm{run}_{t}\big[\mathbf{1}\{V^{\mathrm{AC},m}_{i,t}\notin[\underline{V},\overline{V}]\}]\big]\geq H\Big\},
\label{eq:Ipers}\\
\hat{p}_{\mathrm{pers}}
&=
\frac{1}{M|\mathcal{N}|}\sum_{m=1}^{M}\sum_{i\in\mathcal{N}} I_{\mathrm{pers}}(m,i).
\label{eq:ppers}
\end{align}
\end{subequations}
Undervoltage and overvoltage persistence are reported separately because their remedies differ.

\section{Proposed Risk-Budgeted Outer Loop}
\label{sec:outer}

For a finite candidate set $\Theta$, each $\theta\in\Theta$ generates a fixed envelope $d^{\mathrm{fair}}(\theta)$. A tuning ensemble estimates curtailment, export, voltage risk, and persistence. The selected parameter solves
\begin{equation}\label{eq:calibration_obj}
\min_{\theta\in\Theta}\;
\hat{\mathbb{E}}\big[E_{\mathrm{curt}}(\theta)\big]
+\lambda_{R}\,\mathrm{CVaR}_{\alpha}\big(L(\theta)\big)
\end{equation}
subject to
\begin{equation}\label{eq:calibration_constraints}
\hat{p}_{\mathrm{viol}}(\theta)\leq\varepsilon,
\quad
\hat{p}_{\mathrm{pers}}(\theta)\leq\varepsilon_{\mathrm{pers}},
\end{equation}
or equivalently maximizes expected export. To limit selection bias, feasibility is checked with an upper confidence bound on each empirical risk, and the chosen $\theta$ is re-evaluated on an independent test ensemble. Because $\Theta$ is small, grid search is transparent and parallelizable. This calibrates envelopes under risk budgets; it is not a formal chance-constrained guarantee unless stronger sampling assumptions are enforced.

\rev{In the reported robust instantiation, the empirical objective is replaced by the worst-case expected regime loss $\max_{q:\,\frac12\lVert q-\pi_0\rVert_1\leq 0.20}\sum_{k=1}^{6}q_k\widehat L_k(\theta)$, where $\pi_0$ is the reference regime law, $\widehat L_k$ is the estimated regime loss including curtailment and emergency restoration, and $q=(q_k)_{k=1}^{6}$ is a probability mass function over the six regimes (i.e., an adversarially reweighted mixture). The confidence-bound voltage and persistence constraints remain active.}

% \begin{algorithm}[!t]
%   \caption{Risk-Budgeted Fair-DOE Tuning}
%   \label{algorithm:fair-doe-tuning}
%   \footnotesize
%   \begin{algorithmic}[1]
%     \STATE \textbf{Input:} $\Theta$, $d^{\rm tech}$, tuning/test samples, $\varepsilon$, $\varepsilon_{\rm pers}$, $\alpha$, $\lambda_R$.
%     \STATE Construct each $d^{\rm fair}(\theta)$ and run Algorithm~\ref{algorithm:cvar-recourse} on common tuning samples.
%     \STATE Reject candidates violating a risk upper bound or the emergency-recourse criterion.
%     \STATE Select the feasible candidate minimizing~\eqref{eq:calibration_obj} (or maximizing export).
%     \STATE Freeze $d^{\rm fair}(\theta^*)$ and report an independent test evaluation.
%   \end{algorithmic}
% \end{algorithm}

\begin{algorithm}[t]
  \caption{Risk-Budgeted Fair-DOE Tuning}
  \label{algorithm:fair-doe-tuning}
  \footnotesize
  \begin{algorithmic}[1]
    % \STATE \textbf{Input:} candidate set $\Theta$; technical DOE $d^{\mathrm{tech}}$; tuning and test samples; $\varepsilon$, $\varepsilon_{\mathrm{pers}}$, $\alpha$, and $\lambda_{R}$.
    \STATE \textbf{Input:} candidate set $\Theta$; technical DOE $d^{\mathrm{tech}}$; tuning and test samples; $\varepsilon$, $\varepsilon_{\mathrm{pers}}$; reference law $\pi_0$ and TV radius $0.20$.
    \STATE \textbf{Output:} selected $\theta^{*}$ and independently tested DOE $d^{\mathrm{fair}}(\theta^{*})$.
    \STATE For each $\theta\in\Theta$, construct $d^{\mathrm{fair}}(\theta)$ subject to $0\leq d^{\mathrm{fair}}\leq d^{\mathrm{tech}}$ and the fairness/efficiency constraints.
    % \STATE Run Algorithm~\ref{algorithm:cvar-recourse} on the common tuning ensemble and evaluate~\eqref{eq:calibration_obj}, risk upper bounds, and strict-feasibility failures.
    % \STATE Discard $\theta$ when the upper bound on $\hat{p}_{\mathrm{viol}}$ exceeds $\varepsilon$, the upper bound on $\hat{p}_{\mathrm{pers}}$ exceeds $\varepsilon_{\mathrm{pers}}$, or emergency recourse is unacceptable.
    % \STATE Select $\theta^{*}$ with the smallest mean-plus-CVaR objective among the remaining candidates (or the largest expected export).
    \STATE Run Algorithm~\ref{algorithm:cvar-recourse} on the common tuning ensemble; record regime losses $\widehat L_k(\theta)$, risk upper bounds, and strict-feasibility failures.
    \STATE Discard $\theta$ when a voltage or persistence upper bound exceeds $\varepsilon$ or $\varepsilon_{\mathrm{pers}}$.
    \STATE Select $\theta^{*}$ minimizing the worst-case loss $\max_{q}\sum_{k}q_k\widehat L_k(\theta)$ over $\tfrac12\lVert q-\pi_0\rVert_1\leq 0.20$ (mean-plus-CVaR~\eqref{eq:calibration_obj} is the nonrobust alternative).
    \STATE Freeze $d^{\mathrm{fair}}(\theta^{*})$ and re-run Algorithm~\ref{algorithm:cvar-recourse} on the independent test ensemble; report both tuning and test statistics.
  \end{algorithmic}
\end{algorithm}

\rev{Algorithm~\ref{algorithm:fair-doe-tuning} evaluates $\Theta=\{0.60,0.65,0.70,0.75,0.80\}$ over all six regimes within a total-variation ambiguity radius of $0.20$. All five candidates stay feasible (zero training failures; identical emergency energy), so none is rejected. The worst-case training loss decreases from $42.266$ at $\beta=0.60$ to $40.300$ at $\beta=0.80$, yielding $\beta^\star=0.80$. The selected envelope is then frozen and evaluated over a single test seed, with 1000 samples per probability law.}%with test seed $2026092802$ on 1000 samples per probability law.}
% \rev{Algorithm~\ref{algorithm:fair-doe-tuning} evaluates $\Theta=\{0.60,0.65,0.70,0.75,0.80\}$ over all six regimes within a total-variation ambiguity radius of $0.20$. All candidates are feasible with zero training failures. The worst-case training loss decreases from $42.266$ at $\beta=0.60$ to $40.300$ at $\beta=0.80$, yielding $\beta^\star=0.80$. The selected envelope is then frozen and evaluated over a single test seed, with 1000 samples per probability law.}%with test seed $2026092802$ on 1000 samples per probability law.}

\section{Case Study}

The base case is the IEEE~33-bus radial feeder over 24 hourly intervals with 15~PV units, 8~wind units, and 8~BESSs. The installed renewable capacity is 4.8~MW PV plus 1.9~MW wind; the BESS fleet totals 2.4~MW/7.2~MWh. The original network study used $0.90$--$1.05$~p.u.\ as its hard voltage range and $\beta_t=0.70$ from hours~10--16. Nine regions are coordinated by a distributed OPF solved using ADMM. AC validation fixes the optimized injections and recomposes substation exchange, including nonlinear losses. \rev{Because the risk layer is evaluated only on the balanced IEEE~33-bus feeder already used in~\cite{quessongo2026}, the present study provides internal validation but not cross-feeder generalization; an unbalanced IEEE~123-node or utility feeder is required as an external test.}

\begin{table}[!t]
\caption{Minimal experimental matrix}
\label{table:exp-matrix}
\centering
\resizebox{\columnwidth}{!}{%
\begin{tabular}{lcc}
\hline
\textbf{Experiment} & \textbf{Factors} & \textbf{Outputs} \\ \hline
Mixture shift & Four laws (Table~\ref{tab:probability}) & Risk, severity, persistence \\
Envelope type & Fair vs.\ technical (same $\pi$) & Curtailment, Jain/Gini, risk \\
\rev{Noise ablation} & \rev{Regimes vs. nominal $\xi$} & \rev{Uniform development ($N=600$)} \\
\rev{Robust tuning/test} & \rev{$\beta\in\{.60,.65,.70,.75,.80\}$} & \rev{$\beta^\star=.80$; test $N=1000$} \\ \hline
\end{tabular}}
\end{table}

Table~\ref{table:exp-matrix} summarizes the experiments. \rev{The paired developmental ablation uses 600 uniform samples per DOE/noise setting. The final independent test uses 1000 samples for each reported probability law, with runtime, failures, emergency recourse, fallback, fairness, and physical risk reported jointly.}

\section{Results and Discussion}
\label{sec:results}

\subsection{Regime-Only Baseline Results and Ensemble Reporting}
The existing six-regime runs in~\cite{quessongo2026} provide a deterministic baseline before residual-noise sampling. ADMM converged in $208$--$234$ iterations. Curtailment was largest in the high-PV/low-load regime ($17.528$~MWh), while the Jain index remained at least $0.9817$ in renewable-curtailment regimes. Peak-load/low-renewable and night-peak regimes required $0.0807$ and $0.0137$~MWh of emergency demand response, respectively, and therefore were not strictly feasible without additional flexibility. AC model deviations were below $0.01$~p.u.\ and no voltage or thermal violations occurred under the adopted $0.90$--$1.05$~p.u.\ band at the six representative points. These zero counts must not be generalized to unsampled residual errors.

\begin{table}[!t]
\caption{Verified six-regime baseline from~\cite{quessongo2026}}
\label{table:clagtee-baseline}
\centering
\resizebox{\columnwidth}{!}{%
\begin{tabular}{lccccc}
\hline
Regime & Curt.\ [MWh] & Jain & Gini & max $\Delta V$ & Strict \\ \hline
$\omega_1$ nominal & 5.763  & .9985 & .1244 & .0082 & Yes \\
$\omega_2$ high PV & 17.528 & .9920 & .1096 & .0055 & Yes \\
$\omega_3$ high wind & 12.177 & .9817 & .4216 & .0076 & Yes \\
$\omega_4$ peak/low RES & .071   & 1.0000 & .8910 & .0069 & No \\
$\omega_5$ night peak & .000   & 1.0000 & .1173 & .0062 & No \\
$\omega_6$ mixed & 9.898  & .9965 & .1697 & .0048 & Yes \\ \hline
\end{tabular}}
\end{table}

\begin{figure}[!t]
    \centering
    \includegraphics[width=\columnwidth]{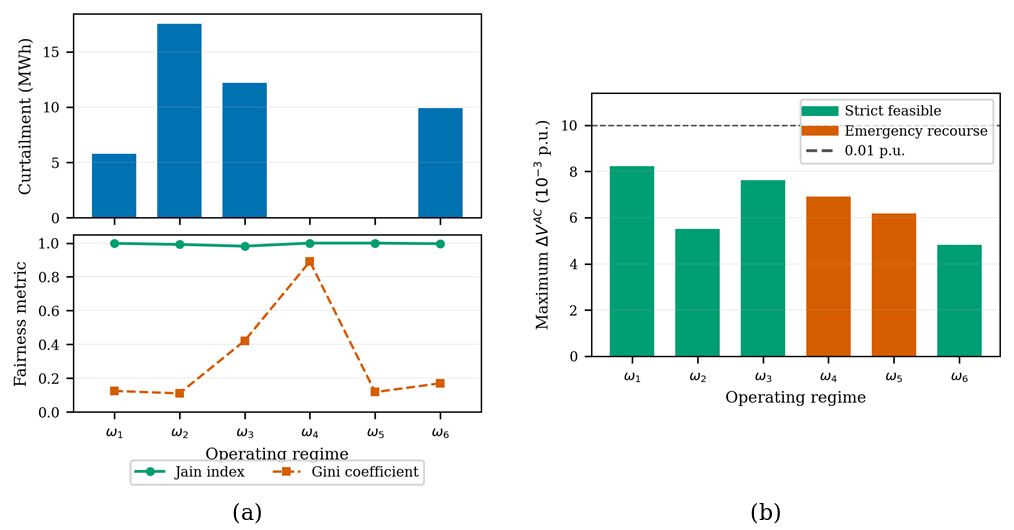}
    \caption{Regime-dependent fairness and AC-validation results:
    (a)~renewable curtailment and ex-post Jain and Gini fairness metrics;
    (b)~maximum AC-versus-LinDistFlow voltage deviation, where orange bars
    identify regimes requiring emergency recourse and therefore lacking
    strict recourse feasibility.}
    \label{fig:fairness_ac_validation}
\end{figure}

Table~\ref{table:clagtee-baseline} presents the deterministic results from~\cite{quessongo2026}. The high-PV regime produces the greatest curtailment, while all regimes maintain high Jain indices and small voltage deviations. However, $\omega_4$ and $\omega_5$ are not strictly feasible because they require emergency recourse. Fig.~\ref{fig:fairness_ac_validation} shows the same trade-off graphically: renewable-rich regimes produce the greatest curtailment while maintaining high Jain indices, the Gini index reveals unequal curtailment in some regimes, and orange bars mark emergency-recourse failures.

{\color{black}
\subsection{AC-Ensemble, Residual-Noise, and DOE-Type Results}

Table~\ref{tab:ensemble-results} reports the independent test of the selected robust DOE. Both probability laws contain 1000 successful nominal-noise samples. All optimizations and AC power flows succeed, and no voltage or persistence event is observed.

\begin{table}[!t]
\caption{\rev{Independent 1000-Sample Test of the Selected Robust DOE}}
\label{tab:ensemble-results}
\centering
\resizebox{\columnwidth}{!}{%
\begin{tabular}{lrrrrrrrrr}
\hline
$\pi$ & $N$ & $\mathbb{E}[C]$ & CVaR$_{.90}$ & Export & Jain & Gini & Strict fail & $\mathbb{E}[\mathrm{DR}]$ & Runtime \\
 & & \multicolumn{3}{c}{(MWh/day)} & & & (\%) & (MWh/day) & (s) \\ \hline
Uniform & 1000 & 6.564 & 17.183 & 1.366 & .9955 & .2890 & 32.8 & .0185 & 3.688 \\
Adversarial & 1000 & 1.954 & 13.153 & .405 & .9985 & .2980 & 73.2 & .0421 & 5.338 \\ \hline
\end{tabular}}
\begin{flushleft}\footnotesize
$C$ is daily curtailed energy. Both rows use $\beta^\star=0.80$, nominal residual noise, and centralized recourse fallback. Violation, persistence, optimization-failure, and failed-AC counts are zero.
\end{flushleft}
\end{table}

% Under the selected robust DOE, the adversarial law reduces expected curtailment relative to uniform but more than doubles strict infeasibility and emergency restoration. The result again shows that low curtailment is not equivalent to low operational risk. Jain coefficient remains high under both laws, whereas Gini coefficient remains near $0.29$--$0.30$.
Under the selected robust DOE in Table~\ref{tab:ensemble-results}, the adversarial law reduces expected curtailment relative to uniform but more than doubles strict infeasibility ($32.8\%$ to $73.2\%$) and expected emergency restoration. Jain's index remains high under both laws, whereas Gini stays near $0.29$--$0.30$, so equity of acceptance does not remove the adequacy risk exposed by emergency recourse.

Table~\ref{tab:noise-ablation-results} isolates the effect of the residual-noise model $\xi|\omega$ before the robust outer loop is applied. Under the Uniform mixture, moving from regime-only to nominal noise raises expected curtailment and CVaR$_{0.90}$ for both fair and technical DOEs, and increases the fair-DOE Gini from $0.326$ to $0.411$. Technical DOEs remain less curtailing and more equal by Gini. The adversarial fair row is a preliminary $N{=}34$ run and is reported only for completeness; the adversarial technical row ($N{=}600$) shows lower curtailment and CVaR than Uniform, consistent with scarce renewables in $\omega_4$--$\omega_5$, but higher expected emergency demand response.

\begin{table}[!t]
\caption{\rev{Developmental Residual-Noise Ablation}}
\label{tab:noise-ablation-results}
\centering
\resizebox{\columnwidth}{!}{%
\begin{tabular}{llrrrrrrrr}
\hline
Mix & DOE &
$\mathbb{E}[C]_{\rm reg}$ & $\mathbb{E}[C]_{\rm nom}$ &
CVaR$_{\rm reg}$ & CVaR$_{\rm nom}$ &
$G_{\rm reg}$ & $G_{\rm nom}$ & $\mathbb{E}[\mathrm{DR}]_{\rm nom}$ \\
 % & & \multicolumn{4}{c}{(MWh/day)} & & & (MWh/day) \\ \hline
Uniform & Fair & 7.05 & 7.96 & 17.53 & 18.69 & .326 & .411 & .0179 \\
Uniform & Tech. & 5.15 & 5.92 & 15.78 & 16.96 & .181 & .194 & .0179 \\
Adv. & Fair$^{\dagger}$ & 2.48 & 2.31 & 15.07 & 16.13 & .424 & .678 & .0452 \\Adv. & Tech. & 1.80 & 1.49 & 12.96 & 11.72 & .114 & .116 & .0432 \\ \hline
\end{tabular}}
\begin{flushleft}\footnotesize
Nominal residual errors use $(\sigma_L,\sigma_{PV},\sigma_W)=(3\%,8\%,10\%)$ with $N{=}600$ except $^{\dagger}$preliminary $N{=}34$.
% Nominal residual errors use $(\sigma_L,\sigma_{PV},\sigma_W)=(3\%,8\%,10\%)$. %$^{\dagger}$Preliminary result based on 34 samples.
\end{flushleft}
\end{table}

Across Tables~\ref{tab:ensemble-results}--\ref{tab:noise-ablation-results}, zero AC-band violations coexist with substantial strict-infeasibility rates that are restored by emergency demand response (e.g., $37.3\%$/$30.5\%$ on Uniform regime-only/nominal developmental samples, and $78.3\%$ on adversarial regime-only samples). The recorded fallback share is $100\%$ in every exported ensemble row, so these Monte Carlo results characterize the disclosed centralized recourse fallback and AC audit rather than samplewise distributed-ADMM convergence.

% Despite zero AC-band violations, strict recourse is infeasible in $37.3\%$ and $30.5\%$ of the uniform regime-only and nominal samples, respectively, and in $78.3\%$ of the adversarial regime-only samples. Feasibility is restored through emergency demand response. Moreover, the recorded fallback share is $100\%$ in every exported ensemble row. Consequently, these Monte Carlo results characterize the disclosed centralized recourse fallback benchmark and AC audit but are not evidence of samplewise distributed-ADMM convergence.

\subsection{Demonstrated Robust Outer Loop}
Table~\ref{tab:outer-loop-results} reports the five-candidate robust tuning grid that produces the $\beta^\star{=}0.80$ envelope tested in Table~\ref{tab:ensemble-results}. All candidates are feasible with zero training failures. The worst-case loss decreases monotonically, so $\beta^\star=0.80$ is selected before the independent test.
% Table~\ref{tab:outer-loop-results} reports the five-candidate robust tuning grid. All candidates are feasible with zero training failures. The worst-case loss decreases monotonically, so $\beta^\star=0.80$ is selected before the independent test.

\begin{table}[!t]
\caption{\rev{Distributionally Robust Outer-Loop Candidate Grid}}
\label{tab:outer-loop-results}
\centering
\resizebox{\columnwidth}{!}{%
\begin{tabular}{rrrrl}
\hline
$\beta$ & Worst training loss & Mean DR sum & Failures & Decision \\
 & & (MWh) & & \\ \hline
.60 & 42.266 & .1098 & 0 & Feasible \\
.65 & 41.664 & .1098 & 0 & Feasible \\
.70 & 41.121 & .1098 & 0 & Feasible \\
.75 & 40.659 & .1098 & 0 & Feasible \\
\textbf{.80} & \textbf{40.300} & \textbf{.1098} & \textbf{0} & \textbf{Selected} \\ \hline
\end{tabular}}
\end{table}

The independent test underlying Table~\ref{tab:ensemble-results} uses 1000 samples per law. Centralized recourse succeeds for every sample, but the ADMM convergence flag is false throughout ($100\%$ fallback), so the AC statistics must not be read as samplewise distributed-ADMM convergence.
% The final test uses 1000 samples per law. Centralized recourse succeeds for every sample, but the ADMM convergence flag is false throughout, producing a $100\%$ fallback share. The selected DOE therefore has verified AC test statistics, but its ensemble performance must not be attributed to samplewise ADMM convergence.
}

\vspace{-2mm}
\subsection{Mixture Interpretation}
For a regime metric $Y_{\omega}$, mixture reweighting changes its expected value without changing the underlying regime outcome:
\begin{equation}\label{eq:mix}
\mathbb{E}_{\pi}[Y]
=
\sum_{\omega\in\Omega}\pi_{\omega}Y_{\omega},
\end{equation}
Using Table~\ref{tab:probability}, expected curtailment is $7.573$, $11.188$, $4.614$, and $2.298$~MWh/day for the Uniform, Sunny-heavy, Night-peak, and Adversarial laws. The adversarial value is lowest only because its stressed regimes have little renewable availability; those regimes instead increase emergency recourse.

\begin{figure}[!t]
    \centering
    \includegraphics[width=\columnwidth]{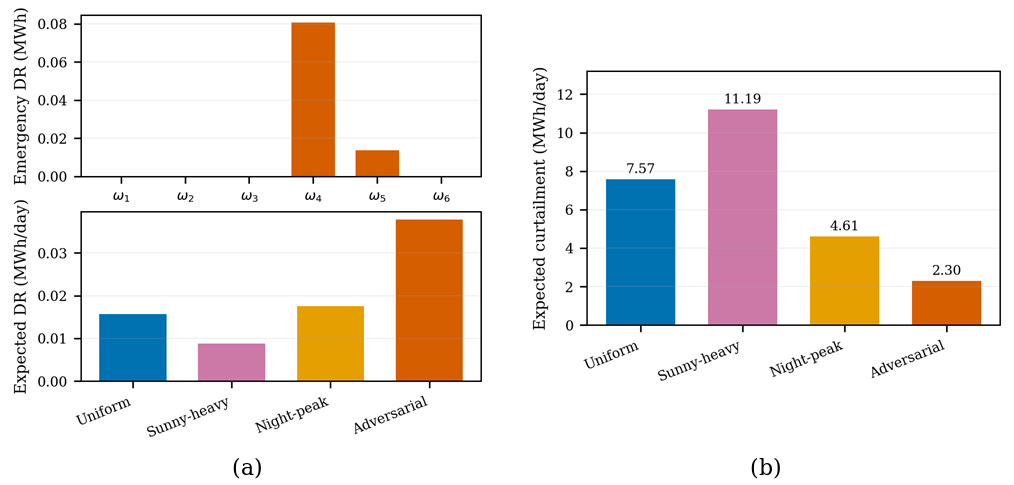}
    \caption{Probability-weighted operational outcomes:
    (a)~emergency demand response required in each operating regime and its
    expected value under the four probability laws; (b)~expected renewable
    curtailment under the same laws. The adversarial distribution produces
    the greatest expected emergency recourse despite exhibiting the lowest
    expected curtailment.}
    \label{fig:expected_recourse_curtailment}
\end{figure}

Fig.~\ref{fig:expected_recourse_curtailment} therefore shows that minimum expected curtailment is not minimum operational risk.

\begin{figure}[!t]
    \centering
    \includegraphics[width=\columnwidth]{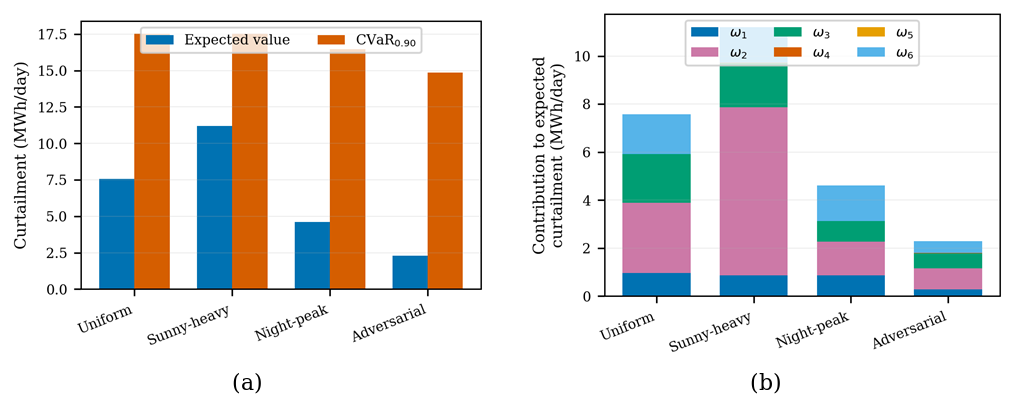}
    \caption{Curtailment risk under the four probability laws:
    (a)~expected daily curtailment and upper-tail $\mathrm{CVaR}_{0.90}$;
    (b)~probability-weighted contribution of each operating regime to
    expected curtailment. The respective regime-only $\mathrm{CVaR}_{0.90}$ values are $17.528$, $17.528$, $16.458$, and $14.853$~MWh/day.}
    \label{fig:cvar_scenario_contributions}
\end{figure}

Fig.~\ref{fig:cvar_scenario_contributions} gives $\mathrm{CVaR}_{0.90}$ values of $17.528$, $17.528$, $16.458$, and $14.853$~MWh/day. The corresponding probability masses on emergency-recourse regimes are $0.333$, $0.150$, $0.550$, and $0.800$, confirming the adversarial law's greater exposure despite lower curtailment.

\begin{figure}[!t]
    \centering
    \includegraphics[width=\columnwidth]{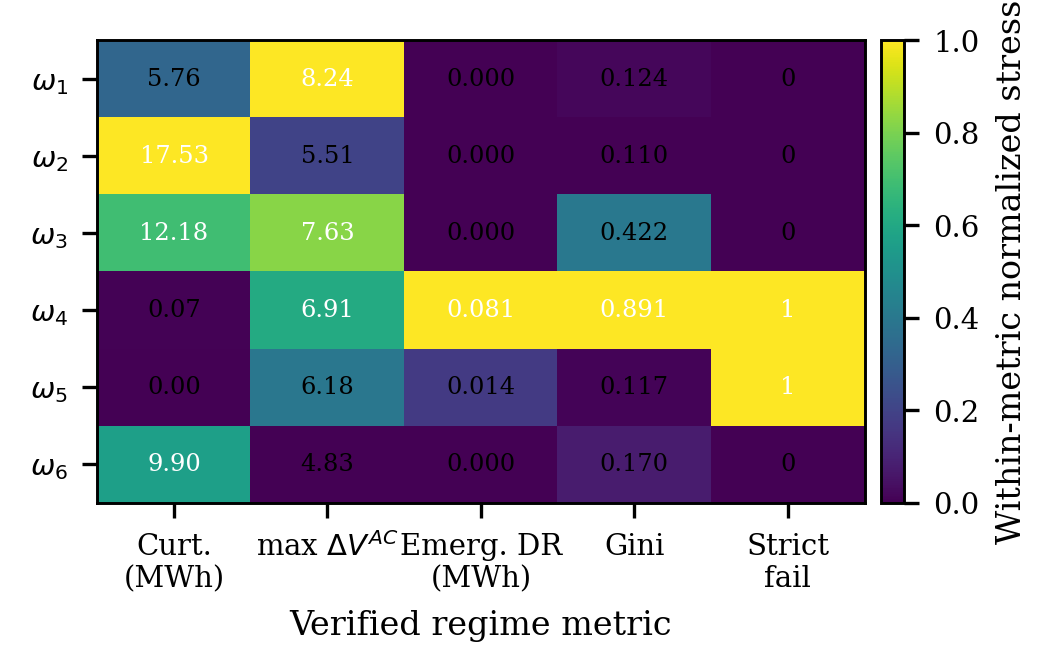}
    \caption{Verified six-regime risk fingerprint. Colours indicate stress
    normalized independently within each metric, whereas the annotations
    report the original physical values. The strict-failure column equals
    one for regimes that require emergency demand response to restore
    feasibility.}
    \label{fig:regime_risk_heatmap}
\end{figure}

Fig.~\ref{fig:regime_risk_heatmap} identifies the most stressed regime per metric and marks emergency-restoration dependence. \rev{Tables~\ref{tab:ensemble-results} and~\ref{tab:noise-ablation-results} extend this regime-centre fingerprint to residual-noise sampling and fair/technical comparison, in which we can verify that AC-band events remain rare, but emergency-restoration rates stay material, so risk cannot be read from curtailment or voltage counts alone.}

% Fig.~\ref{fig:regime_risk_heatmap} identifies the most stressed regime per metric and marks emergency-restoration dependence. \rev{Table~\ref{tab:ensemble-results} complements this deterministic fingerprint,  bus--hour maps are not claimed because the aggregate export does not retain buswise trajectories.}

\vspace{-2mm}
\subsection{Discussion}
No AC-band violation was observed and model deviation remained below $0.01$~p.u.; nevertheless, $\omega_4$ and $\omega_5$ required emergency response, so violation frequency must be read with persistence and strict feasibility. Jain exceeded $0.9817$, yet Gini reached $0.4216$ and $0.8910$ in stressed regimes, confirming that high Jain values can hide concentrated curtailment. \rev{Consistently, Table~\ref{tab:outer-loop-results} selects $\beta^\star=0.80$, and Table~\ref{tab:ensemble-results} shows the corresponding 1000-sample tests with no optimization or AC-flow failures. Still, strict infeasibility of $32.8\%$ (Uniform) and $73.2\%$ (Adversarial) and a $100\%$ fallback share preclude equating these outcomes with samplewise ADMM convergence. Completing sunny-heavy/night-peak ensembles, noise-sensitivity sweeps, and a second feeder remain open.}

% No AC-band violation was observed and model deviation remained below $0.01$~p.u.; nevertheless, $\omega_4$ and $\omega_5$ required emergency response, so violation frequency must be read with persistence and strict feasibility. Jain exceeded $0.9817$, yet Gini reached $0.4216$ and $0.8910$ in stressed regimes, confirming that high Jain values can hide concentrated curtailment. \rev{The robust outer loop selects $\beta^\star=0.80$, and the independent 1000-sample tests complete without optimization or AC-flow failures. However, strict-infeasibility reaches $32.8\%$ under Uniform and $73.2\%$ under Adversarial sampling, while 100\% fallback prevents claims of samplewise ADMM convergence. Sunny-heavy, night-peak, low/high-noise sensitivity, and a second feeder remain open.}

% \vspace{-2mm}
\section{Conclusion}
This paper extended the deterministic fair multiperiod DOE and regional ADMM framework of~\cite{quessongo2026} with an out-of-sample AC risk-assessment layer. The inherited six-regime baseline already shows the interpretive need: high-renewable regimes dominate curtailment, whereas peak/low-renewable regimes can require emergency flexibility despite low curtailment. The proposed common-sample ensemble therefore evaluates fair and technical DOEs jointly through violation frequency, severity, CVaR, persistence, confidence intervals, and equity rather than declaring a DOE fail-safe after a few feasible scenarios. The risk-budgeted outer loop further converts those statistics into a transparent envelope-selection rule. Results remain conditional on the assumed regime probabilities, residual-error model, voltage band, and solver reliability. 

Future work will estimate time-varying regime probabilities from feeder data, extend the model to unbalanced three-phase networks and correlated contingencies, develop distributionally robust risk budgets, and test rolling-horizon recalibration with faster distributed recourse on hardware or utility-data case studies.

% This paper adds an AC-audited risk protocol to fair multiperiod DOEs. \rev{The robust outer loop selects $\beta^\star=0.80$ and the held-out Uniform and Adversarial tests each contain 1000 successful samples. Expected curtailment is $6.564$ and $1.954$~MWh/day, CVaR$_{0.90}$ is $17.183$ and $13.153$~MWh/day, and no voltage or persistence event occurs. Nevertheless, strict infeasibility is $32.8\%$ and $73.2\%$, confirming that minimum curtailment and zero voltage-event counts do not imply minimum operational risk.} Future work will impose strict-feasibility/fallback budgets, complete the remaining sensitivity ensembles, and validate an unbalanced second feeder.

% \vspace{-2mm}
{\fontsize{6.5}{7.2}\selectfont
\bibliographystyle{ieeetr}
\bibliography{references}
}

\end{document}